# A novel fully 3D, microfluidic-oriented, gel-based and low cost stretchable soft sensor

Mohsen Annabestani, Pouria Esmaili-Dokht, Seyyed Ali Olianasab, Nooshin Orouji, Zeinab Alipour, Mohammad Hossein Sayad, Kimia Rajabi, Barbara Mazzolai, and Mehdi Fardmanesh

## Abstract

In this paper a novel fully 3D, microfluidic-oriented, gel-based and low cost highly strechable resistive sensors has been presented. By the proposed method we are able to measure and discriminate all of the stretch, twist, and pressure features by a single sensor which is the potential that we have obtained from the fully 3D structure of our sensor. Against to previous sensors which all have used EGaIn as the conductive material of their sensor we have used low-cost, safe and ubiquates glycol-based gel instead. To show the functionality of proposed sensor some FEM simulations, a set of designed experimental test were done which showed the linear, accurate and durable operation of the proposed sensor. Finally, the sensor was put through its paces on the knee, elbow, and wrist of a female test subject. Also, to evaluate the pressure functionality of the sensor, a fully 3D active foot insole was developed, fabricated, and evaluated. All of the results show promising features for the proposed sensor to be used in real world applications like rehabilitations, wearable devices, soft robotics, smart clothing, gait analysis, AR/VR, etc.

## Introduction

Increasing demand for wearable devices has resulted in the development of soft sensors and actuators. There are some trending wearable devices like smart fabrics [1, 2] and Smart Electro-Clothing Systems [3], sweat sensors, [4, 5], artificial skins [6], health monitoring systems [7-9], and motion capturing devices [10]. Another field that has attracted lots of attention is the employment of soft sensor for the need of virtual reality (VR) and augmented reality (AR) systems [11]. There are a number of applications for this field including immersive entertainment, teleoperation or even physical therapy [12]. Being flexible is one of the consequential factors in this wearable devices. Due to this reason, the field of wearable devices is turning into using soft sensors. There are numerous types of soft sensors including capacitive soft sensors [13, 14], electroactive polymer-based soft sensors [15-18], and resistive soft sensors. Capacitive sensors have shown the ability to detect touch [19-21], stretch [22-24], or touch and stretch without the ability to distinguish between these two [25]. This type of sensor has many advantages but also exhibits nonlinear behavior, they are over sensitive, and some numerical method is needed for identifying the unknown capacitances [26]. The other type of soft sensors is electroactive polymer-based (EAP-based) soft sensors which are divided into two classes depending on the main type of their charge carrier: ionic EAPs [17, 18, 27-29] and electronic EAPs [30]. These sensors despise of having low weight, high sensitivity, biocompatibility, and large produced voltage signal have some drawbacks as well, like being expensive, slow response, sensitivity to moisture and temperature during operation, and their operation is also limited to low temperature due to liquid electrolyte [30]. The third type of soft sensors is resistive soft sensors which are used in this paper. This type has promising features and a better future in comparison with other types. It exhibits a linear response and also can be fabricated using low-cost materials which allows it to be mass-produced.

Various fabrication methods of resistive-based soft sensors have also been proposed. Yong-Lae Park et al. have previously reported the fabrication of a hyperelastic pressure sensing resistive-based device by embedding silicone rubber with microchannels of conductive liquid eutectic gallium–indium [31]. Yong-Lae Park et al. also designed and fabricated a soft artificial skin using the previous method but a different channel design [32]. Mengüç et al. have proposed a design for a soft wearable motion-sensing suit for the application of lower limb biomechanics measurements. The proposed sensors are made of silicone elastomer with embedded microchannels filled with former conductive liquid, EGaIn [33]. Mengüç et al. also use the same method in another paper to design wearable soft sensing suits for human gait measurement [34]. Vogt et al. use the same method and material but exhibits a novel design [35]. In the mentioned paper a soft multi-axis force sensor was proposed that was capable of measuring normal and in-plane shear forces. Joseph T. Muth et al. have reported a new method, known as embedded 3D printing (e-3DP), which involves extruding a viscoelastic ink through a deposition nozzle directly into an elastomeric reservoir. They also used carbon conductive grease, as the functional ink for patterning sensing elements within the 3D printed devices [36]. Seokbeom Kim et al. developed a stretching and twisting sensor that employs a printing method to implement the desired pattern with EGaIn [37].

As it can be seen almost all papers we reviewed previously were using EGaIn as their conductive material which fills the microfluidic channels [38]. EGaIn shows relatively low toxicity, but this cannot be concluded that both ions are non-toxic. Therefore, EGaIn is reasonably safe to use in an aqueous environment, but it should be cautiously handled when any mechanical agitation is applied [39]. On the other hand, EGaIn is a relatively expensive material. To solve this problem, we use conductive gel instead of EGaIn which is non-toxic, durable, and inexpensive. Inspired by reviewed strategies, we proposed a novel method for fabricating resistive soft sensors which enables us to fabricate desired 3D pattern. By making 3D sensors we are able to measure stretch, twist, and pressure at the same time with a single sensor. Yue Zhao et al. also has proposed a method for fabricating conductive 3D metal-rubber composites for stretchable electronic applications [40]. Furthermore Kyung-In Jang et al. introduced a three dimensional network design comprised of helical coils [41]. But these two mentioned methods cannot fabricate any desired pattern. The rest of the paper consist of four parts, in the second part the proposed method is described, in the third part, the fabrications procedure of the proposed soft 3D sensor and the required hardware apparatus will be explained. After that in the fourth part, the results are represented and a discussion on them is done. Finally, in the fifth part, the main points of the proposed sensor are concluded.

## Purposed Method

As stated previously, the simplistic nature of resistive sensors enables us to have a definite prediction of their behavior. The main idea behind the proposed soft sensor is based on measurement the electrical resistance of a conductive material when it's geometry is changing. Theoritaclly when we change the shape of a conductive material, it's electrical resistance will be changed and if the conductive material has geometry with all axis components its resistance will be change in a broder range and can have more functionality facing with external stimulations. Toward these goals first we need a highly soft conductive material and second a 3D structure is needed for conductive material to have all axis components and broder range of sensory functionality. For this aim we have proposed a technique to make 3D microfluidic channels into a highly stretchable materials like silicon rubber to have all axis components. Now If we fill the 3D microchannel with a conductive gel (we call the filled channel as an active part), we will have a soft active part that it has an stable but deformable 3D structure which can response to broder range of mechanical stimulations(Stretch, twist, pressure ,etc) as a highly stretchable soft sensor. In **Fig.1(a)** an example of proposed soft sensor has been illustrated which has tow individual active parts (W-Sensor and L-Sensor) for transverse and longitudinal measurement of mechanical stimulations. We have proposed two other structures which will be described in the next section but in the rest of this section focusing on this structure the basics behind the behavior of proposed soft sensor are investigated.

As our starting point we pick the Pouillet's law and gradually reach a well enough explanation of the sensor's response and finally by a finite element model (FEM) we will show that the proposed idea works and has reliable fidelity to reality.

## Theoretical Justification

As shown in **Fig.1(a)**, designed sensors are made of various compartments in different directions, which is needed for our purposes. For simplicity, the L- sensor, which has long components along the y-axis, has been chosen for the rest of the analysis. Furthermore, we can apply the same principle to the rest of the sensors due to the generalizable argument backing these studies. These sensors' core material is made of a conductive gel that can be interpreted by its resistivity $\rho$. We can divide this sensor into its different elements as $n1$ identical components in the $y$-direction with the length of $L1$ and $n2$ identical components in the $x$-direction with the length of $L2$. Now that we have divided the problem into its components, we can start tackling it more straightforwardly. It can be assumed that $n1$ $y$-direction components and and $n2$ $x$-direction components are connected in series form and so based on the Pouillet's law, the net electrical resistance of all components ($R_T$) of L-Sensor is as follows:

$$R_T = n_1\, \rho_1 \frac{L_1}{A_1} + n_2\, \rho_2 \frac{L_2}{A_2} \qquad (1)$$

In the proposed designs, the thickness and width of the channel are the same throughout the sensor, and the same conductive gel is presented, so $\rho = \rho_1 = \rho_2$ and $A_1 = A_2 = t \times w$, and the following equation will be obtained:

$$R_T = n_1 \rho \frac{L_1}{t \times w} + n_2 \rho \frac{L_2}{t \times w} \quad (2)$$

Where *t* and *w* are thickness and width respectively. We can assumed that for a fixed volume, the $\alpha$ times elongation in any direction leads to an $\alpha$ times cross-section shrinkage and $\sqrt{\alpha}$ times *t* and *w* shrinkage which it can be formulated as follows:

$$Fixed\ Volume = L \times A \rightarrow if\ L' = \alpha L \rightarrow A' = \frac{1}{\alpha} A$$

$$if\ the\ force\ is\ uniform \rightarrow t' = \frac{1}{\sqrt{\alpha}} t\ , w' = \frac{1}{\sqrt{\alpha}} w \quad (3)$$

Any stretch along the y-direction will result in the same situation for the y-oriented components, although there is more complexity for the x-oriented ones. If they were situated at the middle of the sensor, we could assume with confidence that the channel's width will stretch with the same factor, which is not the case here. The amount of width elongation for those channels depends on one more factor: the properties of the flexible material. First, we assume that the width of the x-oriented channels will stretch by a factor of $\beta$ so the equation (2) becomes:

$$R'_T = n_1 \rho \frac{\alpha L_1}{\frac{1}{\alpha} t \times w} + n_2 \rho \frac{\frac{1}{\sqrt{\beta}} L_2}{\beta t \times \frac{1}{\sqrt{\beta}} w} \quad (4)$$

Where $n_2 = n_1 - 1$ and consequently we have :

$$R'_T = \frac{n_1 \rho L_1}{t \times w} \left( \alpha^2 + \frac{1}{\beta} \times \frac{L_2}{L_1} \right) - \frac{1}{\beta} \times \frac{L_2}{L_1} \quad (5)$$

In the specific sensor that is chosen for this analysis (L-Sensor), the length of $L_1$ is far less than of $L_2$ (50 times), and the $\beta$ is near $\alpha$ as stated before, so it is safe to assume those terms negligible and omit them, hence the approximated version of equation (5) is obtained as equation (6) :

$$R'_T = \frac{n_1 \rho L_1}{t \times w} \alpha^2 \quad (6)$$

It is inferred from equation (6) that the output signal of the specific sensor is a 2nd-degree function of the $\alpha$ stretch factor. The same logic can be applied to any other sensors that consist of groups of the same components situated in different orientations. However, it is crucial to consider that not in every sensor can we have the last assumption of negligible length ratios. So the relation will deviate more from the equation (6).

In the last section, we derived a somewhat nonlinear relation between the output resistance and $\alpha$ stretch factor. To test our analysis and complex outputs of the sensors, the dual-stage sensor in **Fig.1(a)** was simulated in COMSOL. To achieve a more realistic response from the simulation, a simple experiment was conducted to determine the relationship between the increased length and the shrunk cross-section. The diagram in **Fig.1(c)** shows how the experimental data is following the theoretical assumption. The mentioned experimental data were used to generate a set of different thicknesses, lengths, and widths and fed to the COMSOL as parameters. Then, in each set of parameters that determined the sensor's mechanical state, resistance was calculated using a 1Khz sinusoidal wave. Finally, the results were fed to MATLAB for post-processing and ratio calculations. It can be seen from **Fig.1(b)** that the mentioned sensors have a second degree behavior which will translate to a more linear characteristic due to the small amount of displacement in real life applications. Although, it is evident that the W-sensor has more nonlinearity due to the more visible nonlinear term in equation (5).

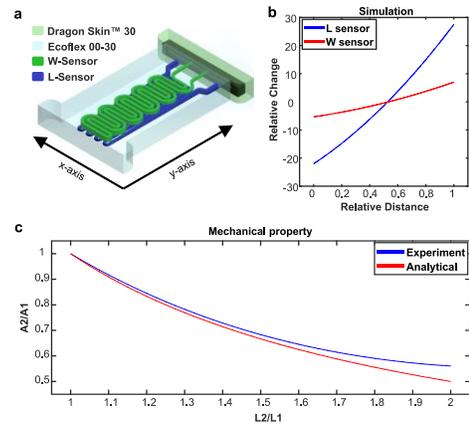

**Fig. 1: sensor design and simulation results (a) Comparison between analytical and experimental data of the mechanical properties (b) sensor design used in the simulations (c) Simulated sensors' characteristics**

At this stage, a sinusoidal and Pulse shape mechanical wave was generated to show the L and W sensor's functionality, shown in **Fig.2**. These examples show that the output electrical signals of the proposed sensors are highly correlated to the input mechanical stimulations and so the sensor can work linearly. In the section four we will validate the response of the proposed sensor by variety of experimental tests.

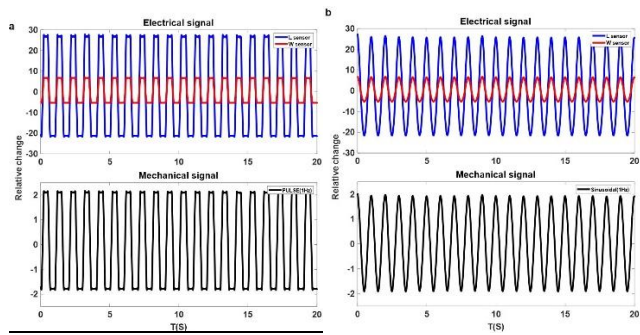

**Fig. 2 : Simulation output of (a) pulse shape and (b) sinusoidal mechanical stimulation**

# Fabrication procedure and Hardware appartus

The fabrication steps for the sensor prototype are represented in **Fig.3**. First, the mold and scaffold are prepared by 3D printing. We used Acrylonitrile Butadiene Styrene (ABS) which is a common thermoplastic polymer as the printing material. Also to prevent the soft material from excessive adhesiveness to the mold, the silicone-oil spray is used as a lubricant, so it can be peeled off easily from the mold. After that, the printed scaffold pattern is inserted into the mold. The Ecoflex 00-30 which is a type of silicon rubber, liquid elastomer, is used as soft material. After preparation of Ecoflex and placing it in the desiccator to expel remaining bubbles, it is poured into the mold. Then it is placed on a hot plate and heated to 120° until Ecoflex is cured. The curing time is about 20 minutes which depends on the thickness of the sensor. After peeling off the chip from the mold, it will be submerged into acetone. By doing this, the acetone vapor dissolves the ABS, and the 3D micro-channels are created into the Ecoflex. In this step, the sensor is ready and can be filled with conductive gel, but because of the high elongation of the Ecoflex, the contact wires may not be fixed in the channel so we have used another material for the clamp region. Thus, after removing ABS from the chip, the clamp region is cut, and the chip is put in the mold again and the clamp region is filled with Dragon Skin™ 30 silicon rubber, again. Dragon Skin™ 30 has lower elongation in comparison with Ecoflex, as the result, this ensures that the wires are well anchored in the soft sensor. After the chip has peeled off to avoid unwanted bubbles into the gel the channels are filled out using a designed syringe pump with the range of mL/min flow rate. Finally, Electrical connections which are made of thin wires will be inserted in each reservoir.

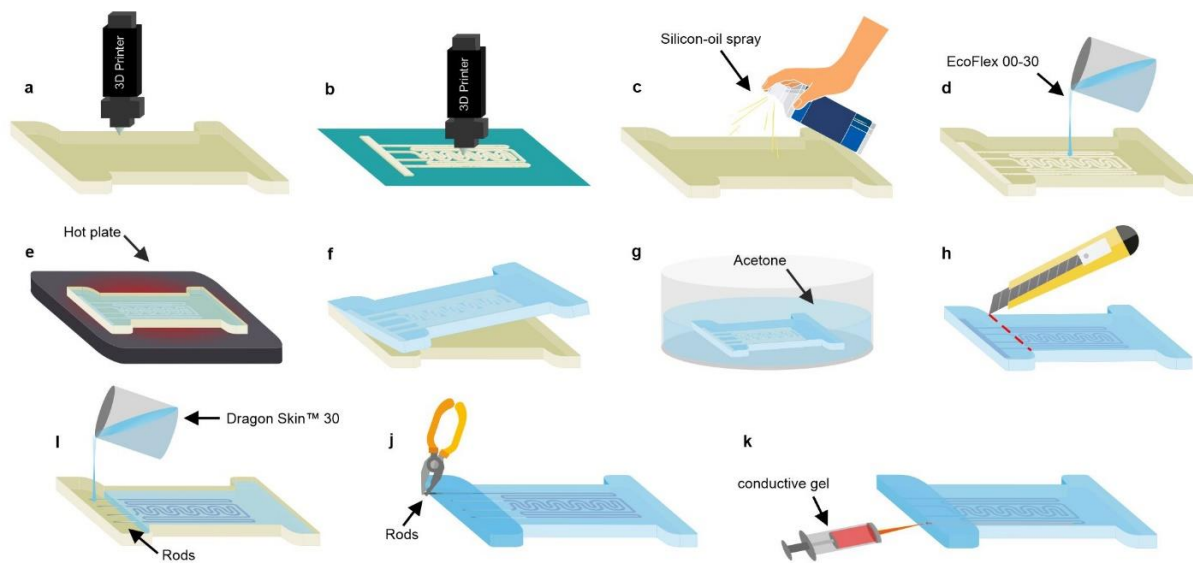

**Fig. 3: Fabrication process (a) 3D printing the mold with ABS. (b) 3D printing the scaffold with ABS. (c) Spraying silicon-oil on the mold (d) pouring the Ecoflex 00-30 into the mold. (e) curing Ecoflex 00-30: heated to 120° for 20 minutes. (f) peeling off the chip from the mold. (g) dipping the chip into the acetone. (h) cropping the clamp region. (i) Filling clamp region with Dragon Skin™ 30 (room-temperature-vulcanizing) silicone rubber. (j) extracting the rods from the chip. (k) filling the channel with conductive gel (combination of water and propylene glycol)**

**Fig.4** shows different sensors for measuring different forces like tension, pressure, and twist. In each of these sensors, we designed two types of sensors. The first one, which we named 'W-Sensor' has maximum sensitivity for transverse force. Another one which we named 'L-Sensor' has maximum sensitivity for Longitudinal force. **Fig.4(a)** illustrates the first sensor which has a planar structure. **Fig.4(b)** shows the second sensor which is designed in a double-stage structure in which the thickness of the channels changes linearly in opposite directions of eachother, means the thickest part of the L-Sensor is where the W-Sensor is in its thinnest part and vice versa. And finally, the third sensor **Fig.4(c)** is designed fully 3D which has a trapozoidal shape in side view.

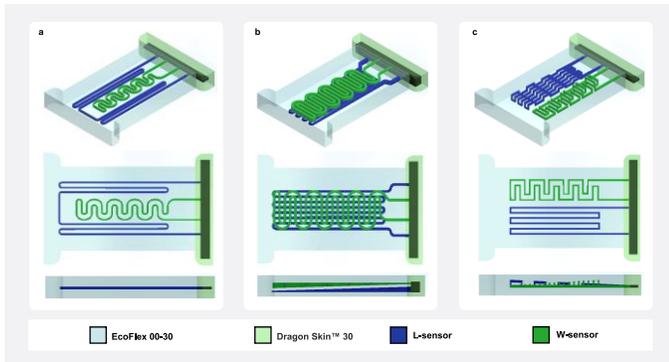

**Fig.4 : Three Sensor designs for different measuring purposes (a) Sensor designed in planar structure (b) Sensor designed in a double-decked structure (c) sensor designed in fully 3D structure**

## Hardware appartus

**Fig. 5(a)** shows the hardware setup flowchart for testing the response of proposed sensors to the applied mechanical signals such as stretch and twist with different waveforms. There is two setup designs for stretching and twisting the sensors, shown in **Fig. 6(a)** and **Fig. 6(b)**, repectively. With the aid of a stepper motor as a mechanical power source, the sensor was stretched or twisted from one stage to another. Furthermore, a camera was used for capturing mechanical stress applied to the sensor using image processing methods, while an electrical setup was logging impedance of the sensor under the test. The flowchart of the whole image processing procedure is shown in **Fig. 5(c)**. For obtaining the exact length of stretch and angle of twist with image processing, we need two black points on each side of the sensor for the stretch setup, shown in **Fig. 6(a)**, and a stick with a black point on its head placed at the end of the sensor for twist setup, shown in **Fig. 6(b)**. In each setup, the captured video will be converted into an image sequence. Then images will turn into BW images by using a threshold that keeps the black points and removes other objects in the picture. In stretch setup, the distance between two black points will be calculated using morphological transforms and other necessary computations. In twist setup, the sensor's twisting angle can be calculated via the position of black point in the camera's view using the same image processing tools. Here we obtained the length of displacement for stretch setup and the angular displacement for twist setup. After that the electrical signal and the mechanical signal were synchronized by an LED, they could be easily coupled, and by use of the coupled signal, the transform function of the following sensor could be obtained as well.

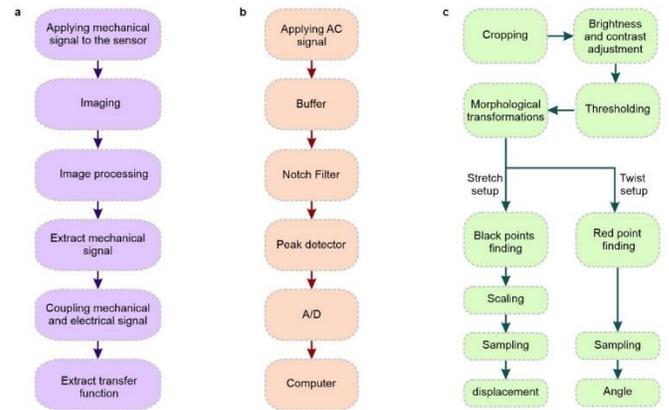

**Fig. 5: flowcharts of sensors' testing setup (a) Depicts the hardware configuration flowchart for measuring the response of proposed sensors to applied mechanical signals such as stretch and twist with different waveforms. (b) A summary of the electrical setup used to obtain and record the sensor's impedance. (c) Flowchart of utilizing image processing techniques to capture mechanical tension applied to the sensor.**

During the experiment, the sensor was subjected to a series of mechanical signals such as Sinusoidal, Ramp, Pulse, and PRBS. All of these triggers were used in both the stretch and twist experiments (**Fig. 6(a)** and **Fig. 6(b)**). The mechanical signals were produced by varying the speed of the stepper motor. **Fig. 5(a)** depicts the flowchart of the electrical setup used to acquire and log the impedance of the sensor. First, an AC signal was applied to an AC voltage divider consisting of the sensor and a Resistor. Then, due to handling the loading issue, the output of the voltage divider was buffered and then was fed to a 50Hz Notch filter to reject city power interference. After that, the output was a clear AC waveform. Because the change in the impedance of the sensor would lead to the change in the amplitude of the AC waveform, a peak detector was used to obtain the amplitude of the signal. In the end, an Analog-to-Digital Converter was used to convert the analog signal to the computer, and then it was sent to the computer by serial communication. An STM32F103 microcontroller which has a 12-bit ADC was used to convert the analog signal to digital and send it to the computer.

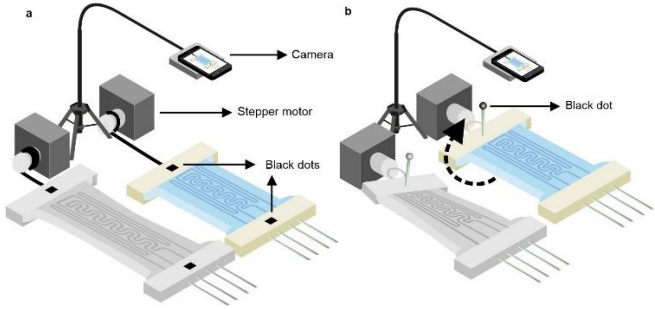

**Fig. 6: The setup designed for logging data from (a) stretching and (b) twisting the sensor.**

## Results and discussion

A series of experiments have been conducted to evaluate the designed sensors and determine their real-world application capabilities. Various mechanical stimulations have been used to achieve a thorough and proper analysis of their behavior, and the variation in the impedance was recorded. Responses to the Sinusoidal and Ramp signals are shown in **Fig. 7(a,b)** and in **Fig. 7(c)**, respectively, indicating excellent performance in the devised circumstances. In order to test the output response to a broad range of frequencies, a mechanical pulse with a fixed frequency and a pseudorandom binary sequence (PRBS) was generated. As shown in **Fig. 7(d,e)**, the electrical signal is effortlessly following the input mechanical stimulation. It can be inferred that these sensors are capable of handling high-speed situations as efficiently as low frequencies. Finally, to discern complex motions and explicate the promising capabilities of these flexible sensors, a more complex mechanical stimulation was devised. In this experiment, sensors were twisted at 90 degrees with the shape of a PRBS wave. As shown in **Fig. 7(f)**, the output signal is well enough explanatory and shows a good following of the input signals. To distinguish the proper behavior of the mentioned sensors in this experiment, more complex data analysis techniques like machine learning and deep learning algorithms are required, which will be investigated thoroughly in future researches.

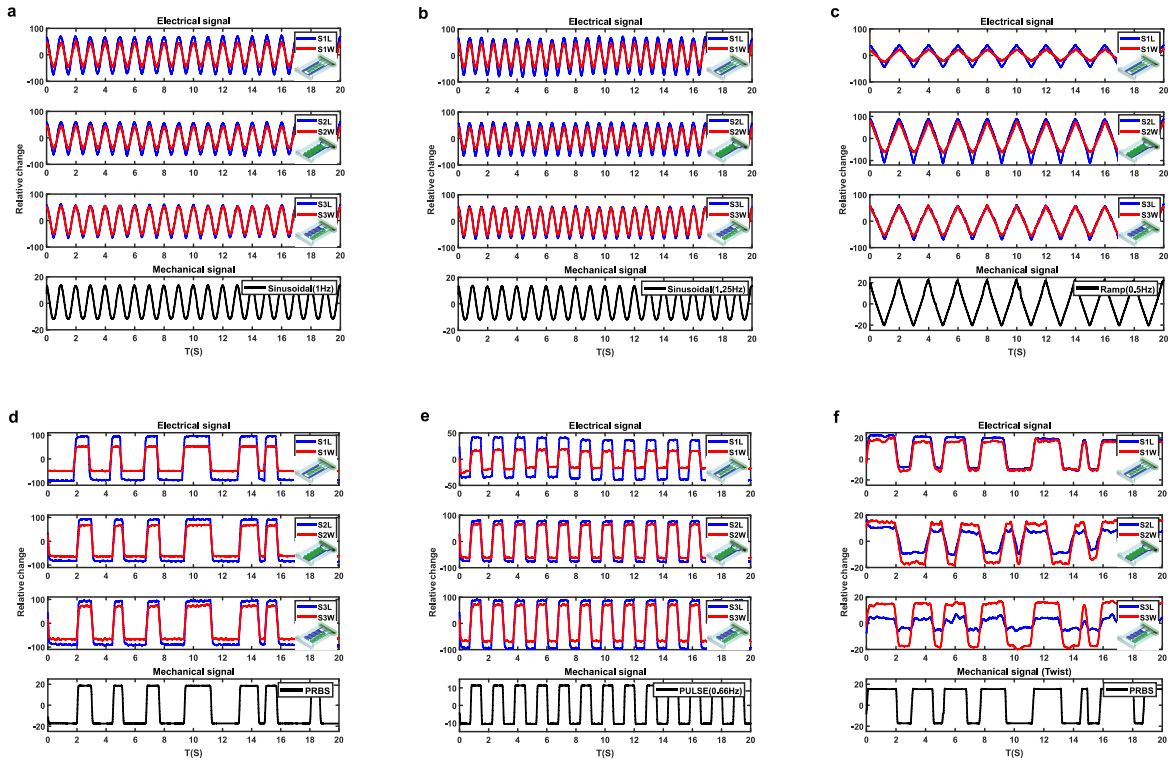

**Fig. 7: The output Experimental results of electrical signal to different mechanical stimulations of (a) and (b) sinusoidal and (c) ramp. (d) and (e) Response of the sensors to the PULSE shape mechanical signals. (f) Response to the twisted mechanical signal.**

## Characterization tests

As stated in the theoretical justification section, the behavior of these flexible sensors is of the second degree with some considerations. Due to the small enough changes along each direction of the sensors in real-world applications, it is expected to have a near-linear behavior that is a massive advantage in data analysis and predictability.

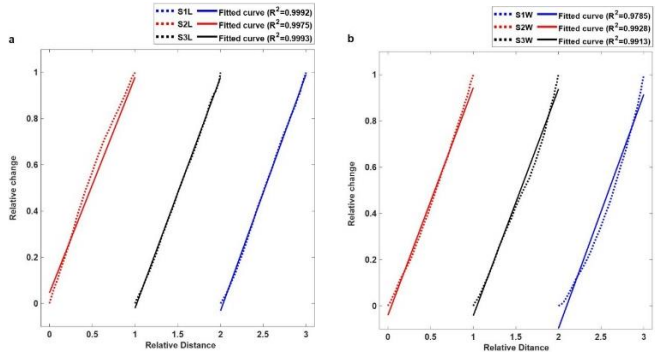

**Fig. 8: The Characterization diagrams of each of the three sensors for both (a) L and (b) W compartments.**

To test this hypothesize, the acquired data in the previous experiments were used to generate a characterization diagram of each of the three main sensors for both L and W compartments, shown in **Fig. 8**. These diagrams and calculated $R^2$ show that the fabricated sensors have excellent linear behavior, which was expected from the results of previous experiments with sinusoidal and ramp signals.

## Durability tests

If we want to use the the sensors in the practical application besides the linearity feature they should also have high level of durability. The linearity of the proposed sensor was investigated in the previous section, here we want to explaine that the proposed sensor is durable enough to be used in real applications. In order to test the durability of our sensor a 4-hour experiment was designed in which the sensor was subjected to stretching back and forth sinusoidally continually in whole of four hours. The results can be seen in **Fig.9** which sensors' performance has not seen a considerable change in this period of time and this fact demonstrates that the purposed sensor is thoroughly durable.

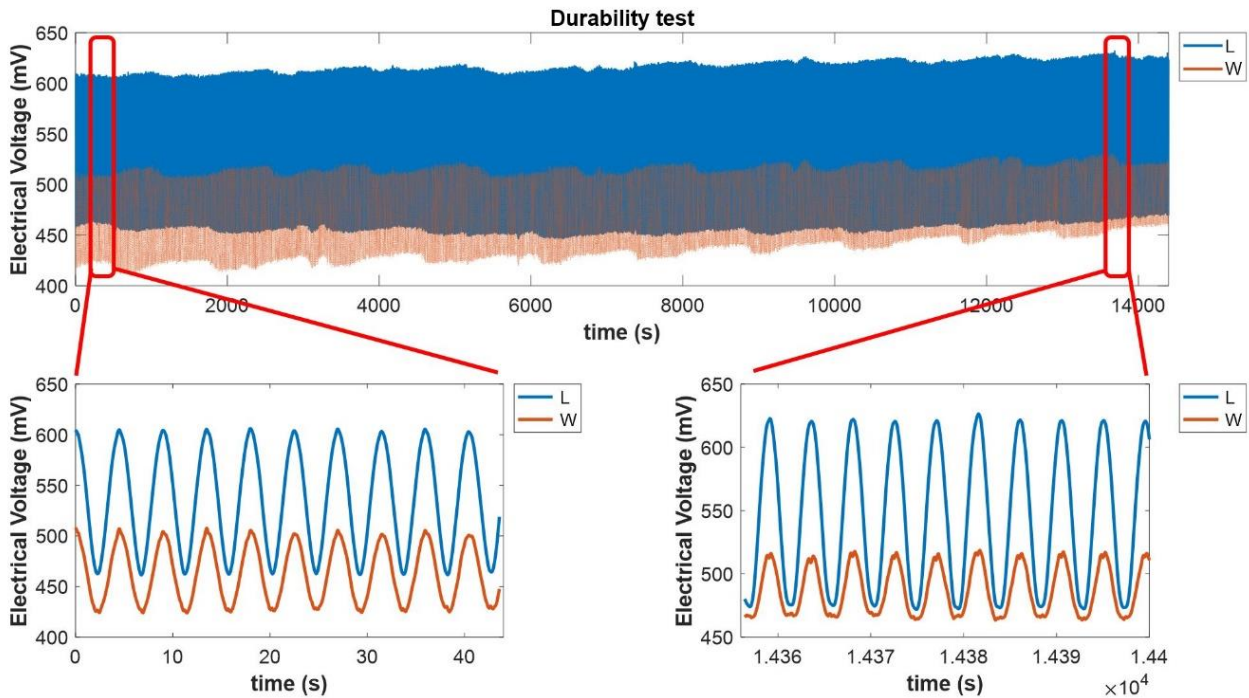

**Fig. 9: Durability test result**

## Real-world applications

Finally, to test the real-world applications of these kinds of sensors, four different experiments were conducted. First, a female subject was asked to integrate sensor number one on her knee and tasked with specific movements, which the results are shown in **Fig.10(a)** and **Fig.10(d1)**. Then the same procedure was used to test the same sensor on a different part of the body as elbow **Fig.10(b)** and **Fig.10(d2)**. The same subject was asked to do specific movements on her elbows to demonstrate another application of these sensors as a measurement system in physical therapies or performance measurements. Finally, to conclude a set of complete movements, the same sensor was placed on the subject's wrist and asked to follow a set of motions which included the twist and bending motion, and the results are shown in **Fig.10(c)**. As you can see in **Fig.10(a,b,c)** the proposed sensor is able to measure the movement of the joints like knee, elbow, wrist etc accurately and also it is abale to detect complex movement like twist of wrist, neck , etc. These results can assure us the proposed sensor has enough potential to be used in real work applications like rehabilitations, wearable devices, soft robotics, smart clothing, gait analysis, motion capture, AR/VR, etc.

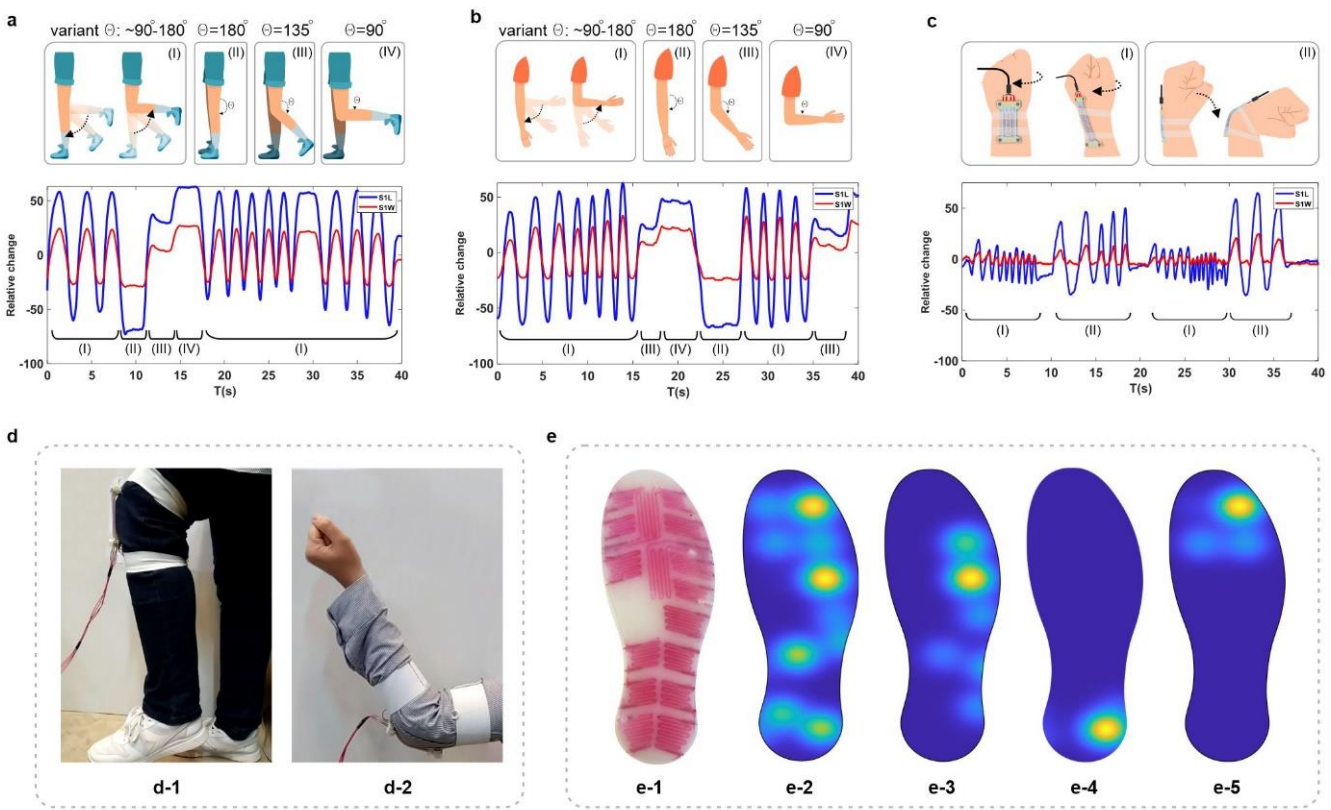

**Fig. 10: Practical results of using the sensor in real-world applications (a)** Output signals of Knee movements such as **(a-l)** moving the knee sinosoidally with variant theta of 180 to 90, **(a-ll)** keeping the knee straight, **(a-lll)** bending the knee at $\theta = 135°$ and also **(a-lV)** bending the knee till $= 90°$. **(b)** Output signals of elbow movements such as **(b-l)** moving the elbow sinosoidally with variant theta of 180 to 90, **(b-ll)** keeping the elbow straight, **(b-lll)** bending the elbow at $\theta = 135°$ and also **(b-lV)** bending the elbow till $\theta = 90°$ **(c)** Output signals of wrist movements including **(c-l)** twist and **(c-ll)** bending motion. **(d)** real images of testing the proposed sensor on subject's **(d-1)** knee and **(d-2)** elbow. **(e)** the whole 3D fabricated foot insole: **(e-1)** real-world image of the insole **(e-2)** the pressure gradient graph as the foot naturally applies pressure on all areas of the insole **(e-3)** the situation where the foot put pressure on side part **(e-4)** heel part **(e-5)** frontal part of the insole.

At last, one of the distinctive features of fabricated sensors was put to the test. The novel fabrication method used in this paper made it possible to fabricate three-dimensional channels that translate to another degree of freedom in sensing applications. As shown in **Fig.4(c)** it is plausible to produce various gradients to grasp a better sense of direction in these sensors and design them for specific purposes. Using **Fig.4(c)** as a reference, a whole active foot insole was fabricated using the same material used as before integrated with sixteen 3D sensors **Fig.10(e1)**. These sensors are able to measure the amount of pressure applied to them due to the 3D structural nature they have. Furthermore, because of the designed gradients in height throughout the foot insole plane, it is feasible to discern the points of pressure, which all are due to the extraordinary capabilities that the new fabrication method brought with itself. The shown diagram in **Fig.10(e2)** until **Fig.10(e5)** explains the excellent pressure pinpoint ability of the active foot insole that can be utilized to screen specific problems such as flat feet and improper walking forms in children. **Fig.10(e2)** shows the pressure gradient graph when the foot puts pressure on all parts of the insole normally. **Fig.10(e3)** depicts the situation where the foot puts more pressure on the side of the insole and also **Fig.10(e4)** illustrates the condition when the foot presses against the ankle. Finally **Fig.10(e5)** shows a posture in which the foot presses against the forefoot.

## Conclusion

In this paper a novel approach for fully 3D, microfluidic-oriented, gel-based and low cost resistive soft sensors was proposed. By the proposed method we are able to quantify and discriminate all of the stretch, twist, and pressure features by a single sensor. To show the functionality of proposed sensor first the basic physics behind this resistive sensor was inevestigated using FEM simulations. Second a set of designed experimental test were done. During the experimental tests, the sensor was subjected to a series of mechanical signals such as Sinusoidal, Ramp, Pulse, and PRBS which all showed the linear and accurate operation of the sensors. Also, During a 4-hour length experiment the durability of the purposed sensor was proofed. Finally, the sensor was put through its paces on the knee, elbow, and wrist of a female test subject. Also, to evaluate the pressure functionality of the sensor, a complete 3D active foot insole was developed and fabricated, and the results were displayed using a gradient graph of the foot.

These results can assure us the proposed sensor has enough potential to be used in real work applications like rehabilitations, wearable devices, soft robotics, smart clothing, gait analysis, motion capture, AR/VR, etc.